\documentclass[twoside,twocolumn,9pt]{article}
\usepackage{extsizes}
\usepackage[super,sort&compress,comma]{natbib} 
\usepackage[left=1.5cm, right=1.5cm, top=1.785cm, bottom=2.0cm]{geometry}
\usepackage{balance}
\usepackage{mathptmx}
\usepackage{sectsty}
\usepackage{graphicx} 
\usepackage{lastpage}
\usepackage[format=plain,justification=justified,singlelinecheck=false,font={stretch=1.125,small,sf},labelfont=bf,labelsep=space]{caption}
\usepackage{float}
\usepackage{fancyhdr}
\usepackage{fnpos}
\usepackage[english]{babel}
\addto{\captionsenglish}{%

}
\usepackage{array}
\usepackage{droidsans}
\usepackage{charter}
\usepackage[T1]{fontenc}
\usepackage[usenames,dvipsnames]{xcolor}
\usepackage{setspace}
\usepackage[compact]{titlesec}
\usepackage{hyperref}

\usepackage{epstopdf}

\definecolor{cream}{RGB}{222,217,201}

\usepackage{bm}
\usepackage{comment}
\usepackage{amssymb,amsfonts,amsmath}
\usepackage{array}
\usepackage{caption, subcaption} 

\begin{document}

\pagestyle{fancy}
\thispagestyle{plain}
\fancypagestyle{plain}{
\renewcommand{\headrulewidth}{0pt}
}

\newcommand*{\fznote}[1]{{\color{blue}({\bf \footnotesize FZ: } #1)}}
\newcommand*{\iknote}[1]{{\color{magenta}({\bf \footnotesize IK: } #1)}}
\newcommand*{\aanote}[1]{{\color{Orange!90!Black}({\bf \footnotesize AA: } #1)}}
\newcommand{\matr}[1]{\boldsymbol{\bm{#1}}}
\renewcommand{\vec}{\bm}
\newcommand*{\rev}[1]{{\color{red}#1}}

\makeFNbottom
\makeatletter
\renewcommand\LARGE{\@setfontsize\LARGE{15pt}{17}}
\renewcommand\Large{\@setfontsize\Large{12pt}{14}}
\renewcommand\large{\@setfontsize\large{10pt}{12}}
\renewcommand\footnotesize{\@setfontsize\footnotesize{7pt}{10}}
\makeatother

\renewcommand{\thefootnote}{\fnsymbol{footnote}}
\renewcommand\footnoterule{\vspace*{1pt}%
\color{cream}\hrule width 3.5in height 0.4pt \color{black}\vspace*{5pt}} 
\setcounter{secnumdepth}{5}

\makeatletter 
\renewcommand\@biblabel[1]{#1}            
\renewcommand\@makefntext[1]%
{\noindent\makebox[0pt][r]{\@thefnmark\,}#1}
\makeatother 
\renewcommand{\figurename}{\small{Fig.}~}
\sectionfont{\sffamily\Large}
\subsectionfont{\normalsize}
\subsubsectionfont{\bf}
\setstretch{1.125} 
\setlength{\skip\footins}{0.8cm}
\setlength{\footnotesep}{0.25cm}
\setlength{\jot}{10pt}
\titlespacing*{\section}{0pt}{4pt}{4pt}
\titlespacing*{\subsection}{0pt}{15pt}{1pt}

\fancyfoot{}
\fancyfoot[RO]{\footnotesize{\sffamily{1--\pageref{LastPage} ~\textbar  \hspace{2pt}\thepage}}}
\fancyfoot[LE]{\footnotesize{\sffamily{\thepage~\textbar\hspace{3.45cm} 1--\pageref{LastPage}}}}
\fancyhead{}
\renewcommand{\headrulewidth}{0pt} 
\renewcommand{\footrulewidth}{0pt}
\setlength{\arrayrulewidth}{1pt}
\setlength{\columnsep}{6.5mm}
\setlength\bibsep{1pt}

\makeatletter 
\newlength{\figrulesep} 
\setlength{\figrulesep}{0.5\textfloatsep} 

\newcommand{\topfigrule}{\vspace*{-1pt}%
\noindent{\color{cream}\rule[-\figrulesep]{\columnwidth}{1.5pt}} }

\newcommand{\botfigrule}{\vspace*{-2pt}%
\noindent{\color{cream}\rule[\figrulesep]{\columnwidth}{1.5pt}} }

\newcommand{\dblfigrule}{\vspace*{-1pt}%
\noindent{\color{cream}\rule[-\figrulesep]{\textwidth}{1.5pt}} }

\makeatother

\twocolumn[
  \begin{@twocolumnfalse}
\vspace{1em}
\sffamily
\begin{tabular}{m{4.5cm} p{13.5cm} }

 & \noindent\LARGE{\textbf{Twitching fiberboids in the collective : A table-top, one-dimensional, active particle gas}} \\
\vspace{0.3cm} & \vspace{0.3cm} \\

 & \noindent\large{Alex Arnhold,{$^{a}$} 
 Falko Ziebert\textit{$^{a,b,*}$} 
 and Igor M. Kuli\'{c}\textit{$^{c,d,*}$}} 
 \\

\vspace{0.4cm} & \vspace{0.4cm} \\

 & \noindent\normalsize{
Fiberboids are filaments that self-propel by rolling along a surface due to
a flux normal to that surface.
In the last few years, fibers using several materials and driving mechanisms have been developed 
that are able to roll persistently. Here we show  using thermally driven 
nylon fibers on a slightly curved glass surface, that fiberboids also come
in another state, that we call ``twitching'': the fiber rolls for a short time and 
then switches direction, implementing on large scales a diffusion-like motion.
We rationalize the occurrence of this effect and experimentally quantify this new dynamic state.    
We then analyze ensembles of twitching fiberboids and
by confining the fibers inside a circular arena to allow them to roll on a single axis only, 
we implement a simple, table-top realization of a one-dimensional non-equilibrium active gas. 
We study experimentally active cluster formation,
their non-equilibrium dynamics
and we rationalize the observations by a simple lattice model.
This model takes into account a Viscek-like alignment of the velocity directions 
of the individual particles within clusters to yield good qualitative agreement 
with the experiments.
Overall our work opens the path to study collectives of self-rolling fibers as an interesting new class of active matter.
} \\
\end{tabular}

 \end{@twocolumnfalse} \vspace{0.6cm}

  ]

\renewcommand*\rmdefault{bch}\normalfont\upshape
\rmfamily
\section*{}
\vspace{-1cm}


\footnotetext{\textit{$^{a}$~Institute for Theoretical Physics, Heidelberg University, Philosophenweg 19, 69120 Heidelberg, Germany}}

\footnotetext{\textit{$^{b}$~BioQuant, Heidelberg University, Im Neuenheimer Feld 267, 69120 Heidelberg, Germany; E-mail: f.ziebert@thphys.uni-heidelberg.de}}

\footnotetext{\textit{$^{c}$~ Institut Theorie der Polymere, Leibniz-Institut für Polymerforschung Dresden e.V., Hohe Straße 6, Dresden D-01069, Germany}}
 
\footnotetext{\textit{$^{d}$~Institut Charles Sadron UPR22-CNRS, 67034 Strasbourg, 
France; E-mail: kulic@unistra.fr}}

\footnotetext{\textit{$^{*}$~Corresponding authors}}


\section{Introduction}

 Ensembles of active particles, constituted of objects utilizing internal energy sources or external energy fluxes
to propel themselves or to exert forces are known to have properties very different
from equilibrium particle systems. 
On the single particle level, they display for instance 
complex mean squared displacements\cite{Howse}
and transition behaviors in traps\cite{Stark_trap}, 
while on the ensemble level, where interactions come into play,
they can display motility-induced phase separation\cite{Cates}, 
flocking\cite{Vicsek}
or more complex clustering states and phases.
On the microscopic level many active particle systems have been proposed,
including a plethora of realizations of microswimmers\cite{Elgeti_swimmers}, 
as well as colloidal rollers \cite{Bricard,Aranson_rollers},
while only few examples exist to date on the more easily accessible macroscopic scale,
for instance shaken discs \cite{Dauchot_disks}
or hexbots \cite{Dauchot_hexbots}.
While anisotropy emerges naturally by the self-propulsion itself,
additional anisotropy can come from particle shape,
as for self-propelled rods. Especially interesting is the case where
the anisotropy is different from the propulsion direction, as for   
transversely propelled colloids and fibers\cite{Vutukuri,Prathyusha}. 

On the macroscopic scale, so-called fiberboids have been introduced \cite{Baumann, Bazir}
by placing a fiber with circular cross-section on a surface that experiences a flux,
e.g.~a heat or humidity flux. This flux couples to the strain and induces
bending, while the confinement to the surface (e.g.~by gravity) forces the curvature
to stay in-plane, which leads to a phase shift between the realized and
the drive-induced curvature, inducing a state of dynamic frustration and 
creating a torque that drives the rolling motion\cite{cat_review}.
This overall mechanism was recently also recovered using finite element simulations,
implementing the contact problem of a cylinder on a heated surface \cite{rollingFEM}.
Many examples of driven fibers have been put forward, see e.g.~Ref.\cite{ZEEM_review}
for a review. These span from  early work on a macroscopic PDMS cylinder,
rolling via swelling induced by a droplet of solvent\cite{Ghatak1},
to nylon, starch, PDMS or LCE-based fibers driven
thermally or hygroscopically in a self-sustained fashion\cite{Yu_Cai_fiber,Bazir}
up to using advanced materials implementing the drive photothermally
or via photo-induced switching of azobenezene 
molecules\cite{AhnElastomer, Yu_Lu_coreshellfiber,Choi_Kim_azobenz_helix_fibers}.
Interesting phenomena reported include 
different modes of motion depending on the substrate 
properties\cite{Zhou_Zheng_multimodal}
or using helical fibers to
perform robotic tasks or navigate through a maze\cite{Zhao_maze,robotics_review}.

While previous works on propelled fibers have focused on persistent rolling motion,
we here show that thermally driven fiberboids can also perform
what we call ``twitching motion'', i.e.~intermittent rolling with random switching of
directions. We first explain and quantify this new mode of motion for single fibers.
We then study its interesting collective dynamics in confinement, effectively
constraining the fibers to one dimension. 
Finally, we rationalize the
experimental data by simple lattice models.
We close by a discussion and by displaying two further interesting effects:
rectifying and switching of the twitching motion by passive particles (small cargos) 
and collective motion induced by ``fiber surfers'' (large, extended cargos, effectively coupling several fibers).

\begin{figure}[t!]
	\centering
	\includegraphics[width=.99\linewidth]{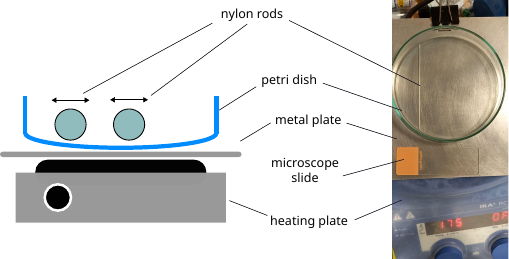}
	\caption{\label{fig_experimental_setup} 
Sketch of the experimental setup, consisting of a heating plate, a metal plate for
homogeneous heating, and a glass container with circular cross-section
as the ``fiber arena'' on top. 
Thermally annealed  nylon fibers (see text; not to scale) are placed inside the arena
at temperatures of $150$ or $175^\circ$C and start to roll. Fiber numbers
between 1 and 16 were studied. The fiber dimension
was adapted such that they can barely rotate their axes in the circular arena, 
allowing for a quasi-one-dimensional dynamics that is recorded with a camera
from above. The microscope slide serves as a scale bar.
		}
\end{figure}

\section{Experiment}
\label{Exp}
 
The experimental setup is sketched in Fig.~\ref{fig_experimental_setup} and
consists of annealed nylon fibers placed in a glass beaker, which is heated from 
below to temperatures in the range $T = {150}^\circ-{175}^\circ$C.
Except for the glass arena, the setup and method is as described in Ref.~\cite{Baumann}:
The nylon fibers are commercially available 
fishing line ("Caperlan $4 \times 4$", supplier \textit{Decathlon}, France) with 
diameter of  $0.6$ mm, cut into pieces 
of a length of  $7.5$ cm. 
As the fishing line as bought is under stresses, 
stemming from the production process, the fibers had  to be annealed first.
Hence, before using them in the experiment, they were placed directly on a metal heating plate,
preheated to ${120}^\circ$C, 
with glass plates  preventing them from falling down.
Over the course of a few minutes, the temperature was 
increased slowly in steps of ${10}^\circ$C until reaching ${175}^\circ$C. 
During the annealing process, the pieces first deformed heavily and then 
straightened out again; 
once temperatures of around ${150}^\circ$C were reached, the fibers 
typically started to roll  in a persistent manner; when the final temperature ${175}^\circ$C was reached, 
the fibers were left for a few minutes after which the annealing procedure 
was considered complete. 
    
\begin{figure}[t!]
    \centering
    \includegraphics[width=.99\linewidth]{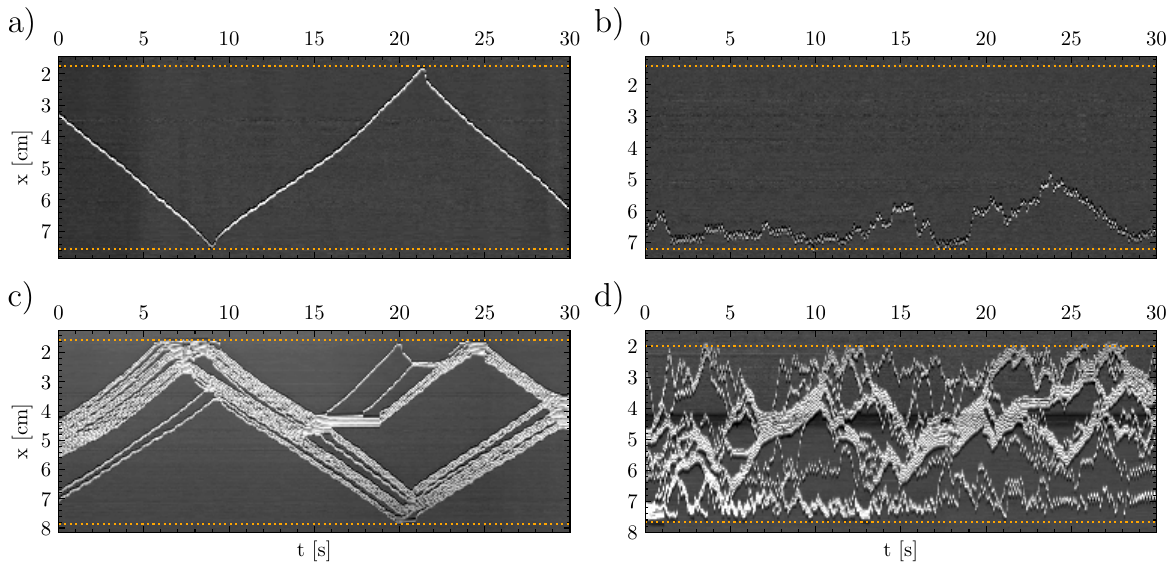}
    \caption{\label{fig_flat_curved_substrate_comparison}
     Effect of flat vs.~curved substrate. a) Kymograph (space-time plot) of a single fiber 
     rolling on a flat glass substrate. The rolling is persistent and the fiber reflects at the
     walls (orange dashed lines on bottom and top).
     b) Same for a single fiber on the slightly curved glass substrate. The twitching motion
     is apparent in the recurrent switches of rolling direction.
     c) Kymograph for the rolling of an ensemble of $N=16$ fibers on a flat substrate.
     d) Kymograph for the rolling of an ensemble of $N=16$ fibers on the slightly curved substrate,
     displaying collective motion and cluster formation on which we focus in this work. 
     Temperature for a)-d): $T = 175^\circ$C. See also supplementary movies 1-4.
    }
\end{figure}

For the data acquisition of rolling fibers,
the heating plate was placed on a leveling lift stand mount
to make sure the experiment was horizontally leveled
and a metal plate was put over the heating plate
to increase the heated region and to
ensure that heating is homogeneous.
On top of all, a glass Petri dish  with circular cross-section of 
inner diameter of ${9.5}$ cm was placed to confine the fibers into
an ``arena''. The fiber length of  $7.5$ cm
was adapted such that they can barely rotate their axes while rolling,
such that the rolling dynamics is quasi-one-dimensional.
Measurements were performed  at a 
"cold" temperature of ${150}^\circ$C and 
a "hot" temperature of ${175}^\circ$C,
recorded via the integrated camera of a \textit{Samsung Galaxy A41} 
mounted on a tripod directly above the Petri dish. 
To gauge the data analysis, a microscope slide of known size
was placed outside of the arena to serve as a scale bar. 
Two different Petri dishes were used:
one with a flat bottom surface, where the behavior was as reported
in Ref.~\cite{Bazir}, i.e.~a persistent rolling of fibers. 
And one with a slightly outwardly curved bottom on which we focus in this work, 
where the rolling was erratic/diffusive, see Fig.~\ref{fig_flat_curved_substrate_comparison} 
for a comparison of the two behaviors.
On both arenas/substrates,
at short time scales the movement of the fibers is one-dimensional 
to very good approximation,
while on longer time scales a precession of the fiber axis and, for multiple fibers,
also a decorrelation of individual axes occur. 
Therefore we analyzed data only up to $60-300$ s, 
to ensure the system stays one-dimensional. 
Fibers were able to explore an ``available arena" region of $\approx 6$ cm,
as shown in the kymographs (space-time plots) 
in Fig.~\ref{fig_flat_curved_substrate_comparison} by the orange dashed lines.

Fig.~\ref{fig_flat_curved_substrate_comparison}a) and b)
show single fibers rolling on the flat and slightly curved glass substrate, respectively.
The former motion is clearly persistent with an approximately constant velocity of $0.4$ cm/s
at  ${175}^\circ$C
and two wall reflections are visible.  
Fig.~\ref{fig_flat_curved_substrate_comparison}b) shows the twitching motion
with frequent reversals of rolling directions. 
The statistics of this motion will be characterized below.
Fig.~\ref{fig_flat_curved_substrate_comparison}c) and d)
illustrate the collective behavior for the two cases:
For the flat substrate, persistent rolling case, the motion stays
persistent and most of the time all fibers roll in the same direction.
In turn, for the curved substrate, twitching case, the dynamics
is much more random: clusters emerge due to partial blocking and
these clusters move collectively with smaller velocities than individual fibers, 
as visible by the reduced slope in the kymograph. Clusters
have only finite ``life times'', as individual fibers leave and join the cluster.
Nevertheless, cluster sizes from two (pairs of fibers) to several fibers coexist at the same time
and the life times of individual clusters can be substantial (seconds).

We here will focus on two main points:
First, to characterize and find the origin of the twitching motion.
And second, to investigate the clustering dynamics occurring in the non-equilibrium, 
driven quasi-one-dimensional system of multiple twitching fibers. 
To analyze the experimental data, fiber tracking was performed 
using the intensity profiles of the kymographs. 
Available software did not yield very satisfactory results, 
hence we implemented a custom made tracking routine 
based on a greedy cost algorithm
that directly implemented the two most important constraints of the dynamics:
namely, that fibers do not change order as they cannot pass, 
and a penalization of overlap of fibers in regions of high intensity,
to discern multiple fibers.

\begin{figure}[t!]
    \centering
    \includegraphics[width=\linewidth]{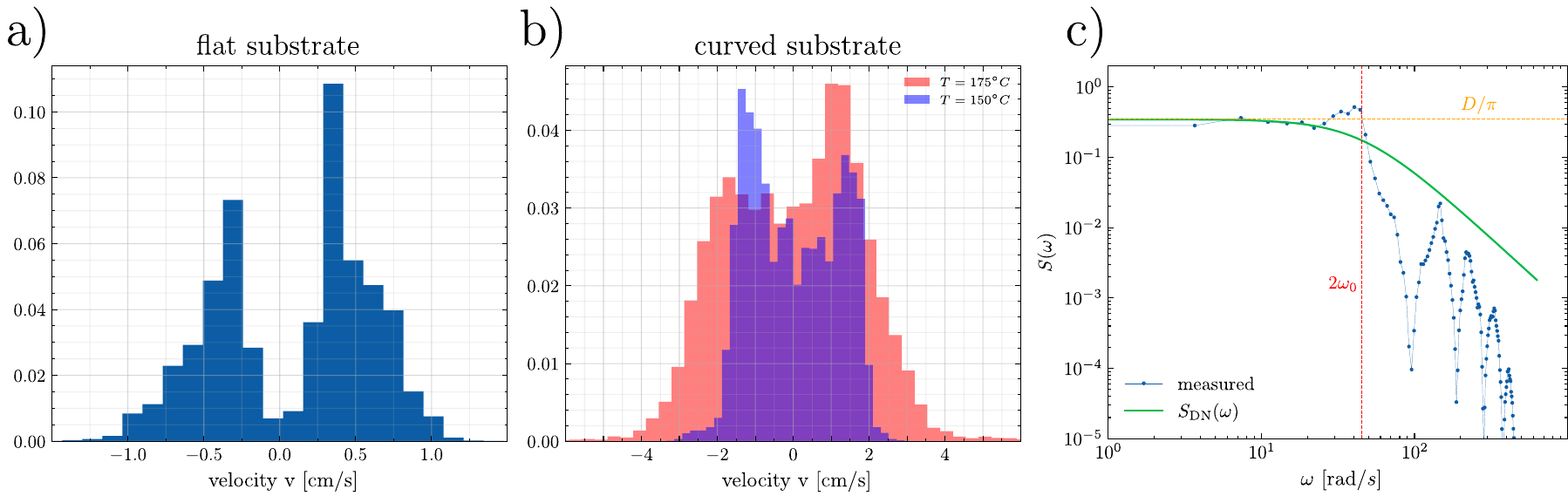}
    \caption{\label{fig_velocity_distribution_flat_curved}
        Distributions of rolling velocities.
        a) A single fiber rolling on a flat substrate at temperature $T = 175^\circ\text{C}$,
        displaying clearly separated peaks at the persistent rolling velocity.
        b) Single twitching fibers on  the slightly curved substrate for driving 
        temperatures $T = 150^\circ\text{C}$ (blue) and $175^\circ\text{C}$ (red). 
        The distribution is flatter with low velocities populated but also broadened.
        c) Power spectral density (PSD) 
        of the velocities for the twitching case at $T = 175^\circ\text{C}$,
        allowing to extract the diffusion coefficient. 
        The green curve shows a fit to the PSD of a perfectly dichotomic process, 
        allowing to also estimate the switching rate between the two rolling directions, 
        verified in addition
        by direct counting of reversals for long trajectories.  
    }
\end{figure}

{\bf Single fiber motion.}
We first investigated the statistics of single fibers.
The obtained statistics for the rolling velocities are shown in 
Fig.~\ref{fig_velocity_distribution_flat_curved}.
Panel a) shows the case of the flat substrate, 
where one sees clear peaks at $\approx \pm 0.4$ cm/s 
reflecting the persistent rolling.  
In contrast, for the curved substrate, twitching motion 
the distribution is much flatter, cf.~panel b). 
While peaks are still discernible, low velocities are highly populated. 
The distribution broadens with temperature, implying larger velocities, 
but the location of peaks at $\pm 1.5$ cm/s depends little on the driving
temperature. The fact that the velocity in the twitching case 
is larger than for the flat substrate (also visible in the kymographs) 
is due to the transient dynamics before steady rolling is established, 
studied in more detail in Ref.~\cite{Bazir}.
Note that similarly, upon a wall reflection, the velocity is first large and than
relaxes to a steady-state value, cf.~Fig.~\ref{fig_flat_curved_substrate_comparison}a).
Finally, Fig.~\ref{fig_velocity_distribution_flat_curved}c) shows 
the power spectral density (PSD) obtained from the velocities,
for the curved substrate case at $T = 175^\circ\text{C}$. 
Its low-frequency (long-time) behavior allows to 
estimate the diffusion coefficient, i.e.~the randomness
of the motion. Assuming the velocities to be perfectly dichotomic, 
we fit the PSD to the result for a dichotomic process\cite{Bena_dicho},
\begin{equation}
S(\omega) = \frac{D}{\pi} \left(1 + \frac{\omega^2}{(2\omega_0)^2}\right)^{-1} ,
\end{equation}
which yields $D = 0.20\pm 0.03{\rm cm}^2/{\rm s}$ and 
a flip-rate (between the two rolling directions) of $\omega_0=3.2\pm0.7{\rm s}^{-1}$.
Because panel b) of Fig.~\ref{fig_velocity_distribution_flat_curved} shows that the process 
is not perfectly dichotomic (it has a broader, continuum velocity spread),
we verified the order of magnitude of the flip-rate also by counting the velocity 
reversals directly from long trajectories,
which yielded $\omega_0 = (7.8 \pm 1.8){\rm s}^{-1}$, in reasonable agreement.

{\bf Origin of the twitching motion.} 
Why is the behavior so much different for the flat vs.~the slightly curved substrate,
as evident from Fig.~\ref{fig_flat_curved_substrate_comparison}? 
To understand this, it is important to note that all fiberboids are slightly curved
when rolling, with a in general time-dependent curvature $\kappa(t)$.
On a flat substrate, after a transient, a steady state
curvature $\kappa_0$ and rolling velocity $v_0$ is reached, similar as in Fig.~\ref{fig_flat_curved_substrate_comparison}a)
after a wall reflection.
In more detail, as shown and explained in Ref.~\cite{Bazir}, the 
thermally induced steady state fiber curvature 
and the fiber's rolling angular frequency $\omega_0=v_0/R$ ($R$ the radius of the fiber's cross section)
fulfill the relation  
\begin{equation}
|\kappa_0|=\frac{p|\omega_0|/R}{\omega_0^2+r^2},
\end{equation}
with $p\propto T$ the thermal pumping rate and $r$ the thermal relaxation rate.
This curvature-velocity relation is non-monotonic
and at the onset of motion, the fiber is close to the branch were the curvature
{\it decreases} with velocity until it reaches its steady state curvature $\kappa_{0}$
within a sub-second time scale (note that in Ref.\cite{Bazir}, fibers were slightly thicker
and the time scale hence about a second). 

Now, if the substrate is slightly curved itself, let us call its curvature $\kappa_s$,
the two curvatures have to negotiate with each other. 
If one places an initially  straight fiber on the slightly curved substrate 
-- and deformation by gravity is weak (i.e.~if the fiber is not very heavy or not very long) --
it will first be ``decoupled'' from the substrate in the absence of thermal effects.
Only when the thermally induced curvature -- both from (partial) thermal contact
and radiative heat transport -- becomes larger than the substrate curvature $\kappa_s$ 
can the fiber embed into the surface and roll.

\begin{figure}[t!]
	\centering
	\includegraphics[width=.9\linewidth]{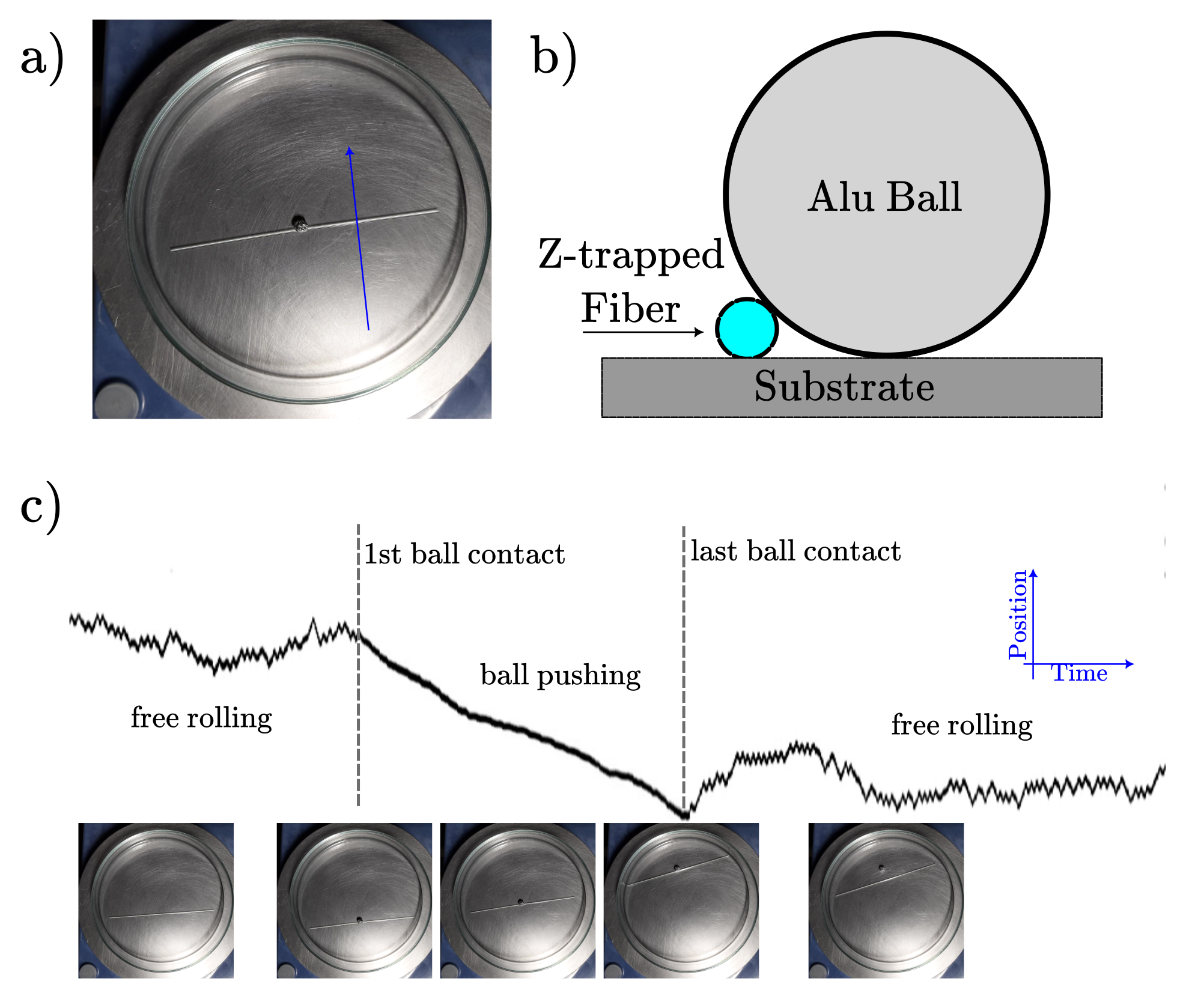}
	\caption{\label{fig_AluFoil-Ball}
	Fiber confinement to the surface triggers its transition from twitching to persistent rolling.  
	a) A fiberboid undergoing twitching motion collides with a millimetric obstacle,
	a ball made of aluminum foil. 
	b) The obstacle, due to its larger size and weight, entraps the fiber in the $z$ direction. 
	c) A kymograph showing the fiber motion together with snapshots at different 
	time points ($x$-axis of kymograph). Upon first contact with the ball, the induced
	confinement triggers a change of fiberboid behavior: it now rolls persistently, 
	pushing the ball ahead of itself. Once the edge of the beaker arena is reached 
	and the obstacle is swept out of way, the fiber resumes random 
	twitching motion once again. Temperature $175^\circ$C. 
	The time span of the kymograph is $29$ s. 
	See supplementary movie 5.    
	}
\end{figure}

This suggests the following scenario with an ever changing curvature of the
fiber $\kappa(t)$: 
(i) A fiber with curvature $\kappa(t_0)<\kappa_s$ (or initially straight) 
is heated and obtains a thermal curvature 
$\kappa(t_1)=\kappa(v_1=0)>\kappa_s$
which allows it to ``embed'' into the surface by falling on either side.
(ii) Then its curvature relaxes as the rolling velocity increases. 
(iii) When it reaches a  curvature $\kappa(t_2)=\kappa(v_2)<\kappa_s$,
it decouples from the substrate again and the process restarts.
At this point, it can stop and possibly change direction (or not)
while its curvature increases again beyond $\kappa_s$.

Since both the substrate and fiber curvatures involved are tiny, our hypothesis
is hard to verify by direct observation within our setup.
However, to test it indirectly, we have placed various obstacles in the way of a twitching fiberboid. 
We found that small mobile obstacles that are curved (a little aluminum foil ball), such as to push down 
and confine the fiberboid stronger to the surface, tend to induce a shift of its behavior 
from twitching to persistent motion, see Fig.~\ref{fig_AluFoil-Ball} 
and  corresponding supplementary movie 5. 
On the other hand, obstacles that do not confine the fiberboid in the $z$ direction 
(such as an aluminum cube; not shown), 
even though they are still moved by the fiberboid, leave the twitching motion unchanged in character. 
While not entirely proving our hypothesis, this finding strongly suggests the
involvement of the ``substrate decoupling'' and  out-of-surface motion 
as a crucial ingredient of the twitching movement.

\begin{figure}[t!]
    \centering
    \includegraphics[width=\linewidth]{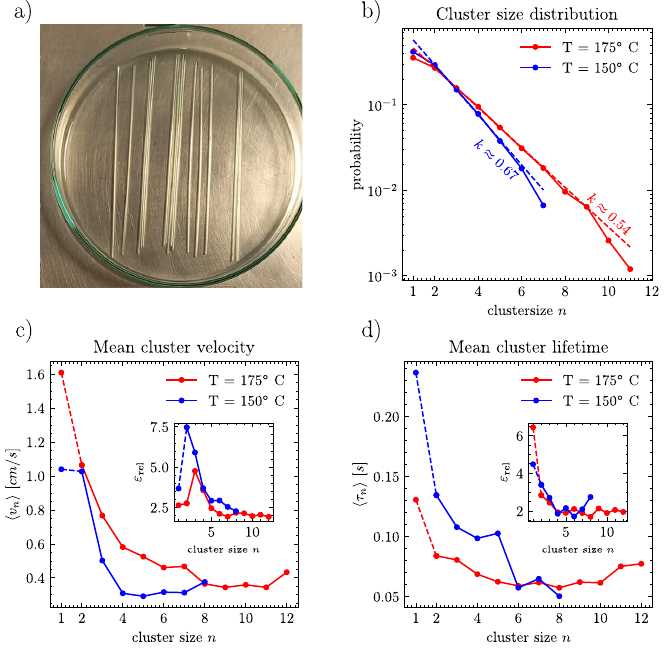}
    \caption{\label{fig_cluster_size}
    Cluster analysis. 
    a) An experimental snapshot showing 15 rolling fibers with 
    a $n=3$ cluster and a $n=5$ cluster and several single fibers.
    b) Semi-log plot of the cluster size distribution, i.e.~the probability to find a cluster 
    of size $n$,  for the two different driving temperatures.
    The fits (to $Ae^{-kn}$) indicate that the probability scales 
    exponentially with size. 
    c) Mean velocity of a cluster of size $n$, for the two different driving temperatures.
    d) Mean lifetime of a cluster of size $n$, for the two different driving temperatures.
     Trajectories of total time $\approx 40-300$ s were analyzed. 
    }
\end{figure}

{\bf Clustering.} We next investigated the clustering of twitching fibers.
Fig.~\ref{fig_cluster_size}a) shows a representative experimental snapshot 
of an example cluster of size $n=3$,
another cluster of size $n=5$ and few single fibers.
The clustering was analyzed for both examined temperatures, 
$T = 150^\circ\text{C}$ (blue curves) and $175^\circ\text{C}$ (red curves).
Fig.~\ref{fig_cluster_size}b) displays the cluster size distribution
via the probability to find a cluster of size $n$. 
Note that the plot is semi-logarithmic,
indicating that the probability to find a cluster decays roughly exponentially 
with cluster size $n$, with temperature dependent exponents as indicated. 
Expectedly, small clusters dominate, but e.g.~$n=4$ clusters have an appreciable
probability of $\sim 10\%$. The hotter, stronger driven system has slightly 
more larger clusters. Note that an exponential distribution of cluster sizes
has been theoretically predicted in Ref.\cite{Soto_Ramin} for run-and-tumble particles
with exclusion, and seems to hold here in good approximation as well.

Second, as also visible in the kymograph of
Fig.~\ref{fig_flat_curved_substrate_comparison}d),
the clusters are not static, but move with a finite -- and size-dependent --
velocity. 
This cluster velocity is quantified in Fig.~\ref{fig_cluster_size}c).
For the larger temperature, which shows a clearer trend,
clusters of size $n=2$ have a velocity reduced by about 30-40\%
compared to single fibers. The velocity of larger clusters levels for 
small drive and continuously decreases
for large drive, but nevertheless stays finite. 
Hence the fibers collectively decide upon one direction
to which they collectively roll, cf.~again Fig.~\ref{fig_flat_curved_substrate_comparison}d).
One may interpret these rolling clusters as a, at least transient, flocking state
occurring especially at the larger temperature.

\begin{figure}[b!]
    \centering
    \includegraphics[width=\linewidth]{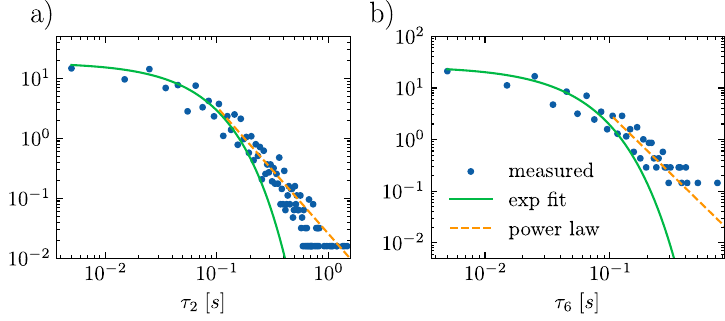}
    \caption{\label{fig_cluster_life}
        The lifetime distribution of clusters of a given size $n$. 
        a) $n=2$ and b) $n=6$.
        The green curves show exponential fits to the data; the orange curves power law fits 
        to the larger times window. The exponents for the power laws are 
        given by $p=-2.1\pm 0.1$ for the cluster of size $n=2$ 
        and $p=-2.4 \pm 0.2$ for $n=6$. 
    }
\end{figure}

Finally, we extracted the mean lifetime $\langle \tau_n\rangle$ of a cluster of size $n$.
The distribution is shown in Fig.~\ref{fig_cluster_size}d), 
evidencing that the mean lifetime decreases with cluster size.
For the low temperature, single fibers
stay isolated for quite some time and large clusters are short lived. 
For the larger temperature, however, larger
clusters have quite comparable lifetimes as $n=2$ clusters. 
The lifetime distribution of a cluster of a given size $n$ is also non-trivial, 
as shown by the histograms in Fig.~\ref{fig_cluster_life}. 
There, clusters of size $n=2$ and $6$ were analyzed and plotted in double-logarithmic scale. 
An exponential distribution (green) is unable to describe the statistics well.
The data rather suggests the lifetime of a cluster of given size 
to be exponentially distributed at early times, crossing over to a power-law tail (orange)
of exponent $\approx -2$ for larger times.

\section{Modeling}
\label{Modeling}

\begin{figure}[t!]
	\centering
	\includegraphics[width=0.99\linewidth]{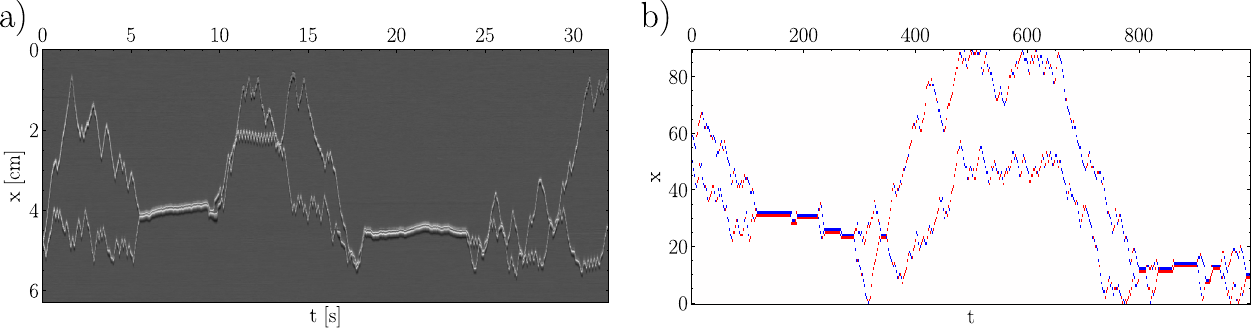}
	\caption{\label{fig_doublet}
	a) Experimental kymograph of a two fiber system,
	displaying long lived ``doublet'' states, where the fibers roll against each other. See also
	supplementary movie 6.
	b) Numerically obtained kymograph for a two fiber system 
	in the ``exclusion model with blocking''.	
	Color-code: red/blue means motion in positive/negative direction
	(i.e.~moving up/down in the figure). 
	The numerical data corresponds to experimental data with rescaled units, 
	where the fiber diameter is $1$ and the fiber velocity $v=1$.	
	Numerical parameters: $\gamma=1$, $\omega_0=0.28$, $\omega_b=0.028$, $L=90$, $t=1000$.
	}
\end{figure}

We now aim to find a minimal model that describes the main features of the experimental system. 
To this end, we first studied binary interactions in more detail. 
Experimentally, for two interacting fibers we observe
states where the fibers roll against each other, forming a ``doublet state'' 
that survives for long times compared to single fiber switching events. 
This doublet then separates again, may form another doublet state etc., 
as shown by the kymograph in Fig.~\ref{fig_doublet}a),
see also supplementary movie 6.

This observation motivated us to introduce what we call 
the ``exclusion model with blocking''.
It is a discrete stochastic lattice model,
with a hopping rate $\gamma$ and a second 
rate $\omega_0$ of switching the direction.
It naturally has to account for excluded volume -- a site occupied 
by a particle cannot accommodate
another particle, cf.~the asymmetric exclusion process \cite{Derrida}.  
Last but not least, to mimic the stubborn ``rolling against each other''
we add another, reduced rate of switching 
the direction, $\omega_b<\omega_0$, in case the site
in the rolling direction is blocked. These rules are schematically sketched in
Fig.~\ref{fig_model_explanation}a)-c), and more details are given in appendix B.
Note that for models of this type (without the additional blocked flip rate),
quite some literature exists, see e.g.~Ref.~\cite{Krapivsky_book}.

This ``exclusion model with blocking'' indeed reproduces the doublet state, 
see the numerical kymograph in Fig.~\ref{fig_doublet}b).  
However, it does not correctly predict the properties 
of larger collectives (mean cluster size, life time and cluster velocity) 
as shown for the cluster size distribution below in Fig.~\ref{fig_model_results}a).
This suggests a missing ingredient in the dynamic rules of the model, 
and also that the behavior of $n=2$ clusters is different from larger clusters.
This can be rationalized by noting that, upon formation,  
$n=2$ clusters completely block each other at first, as they were colliding head on
symmetrically.
This is different for clusters with $n>2$: a $n=3$ cluster 
is asymmetric and larger clusters (even of an even number such as $n=4$) typically emerge
when a $n-1$ cluster is joined by an additional fiber and hence are typically also asymmetric.

\begin{figure}[t!]
    \centering
    \includegraphics[width=0.99\linewidth]{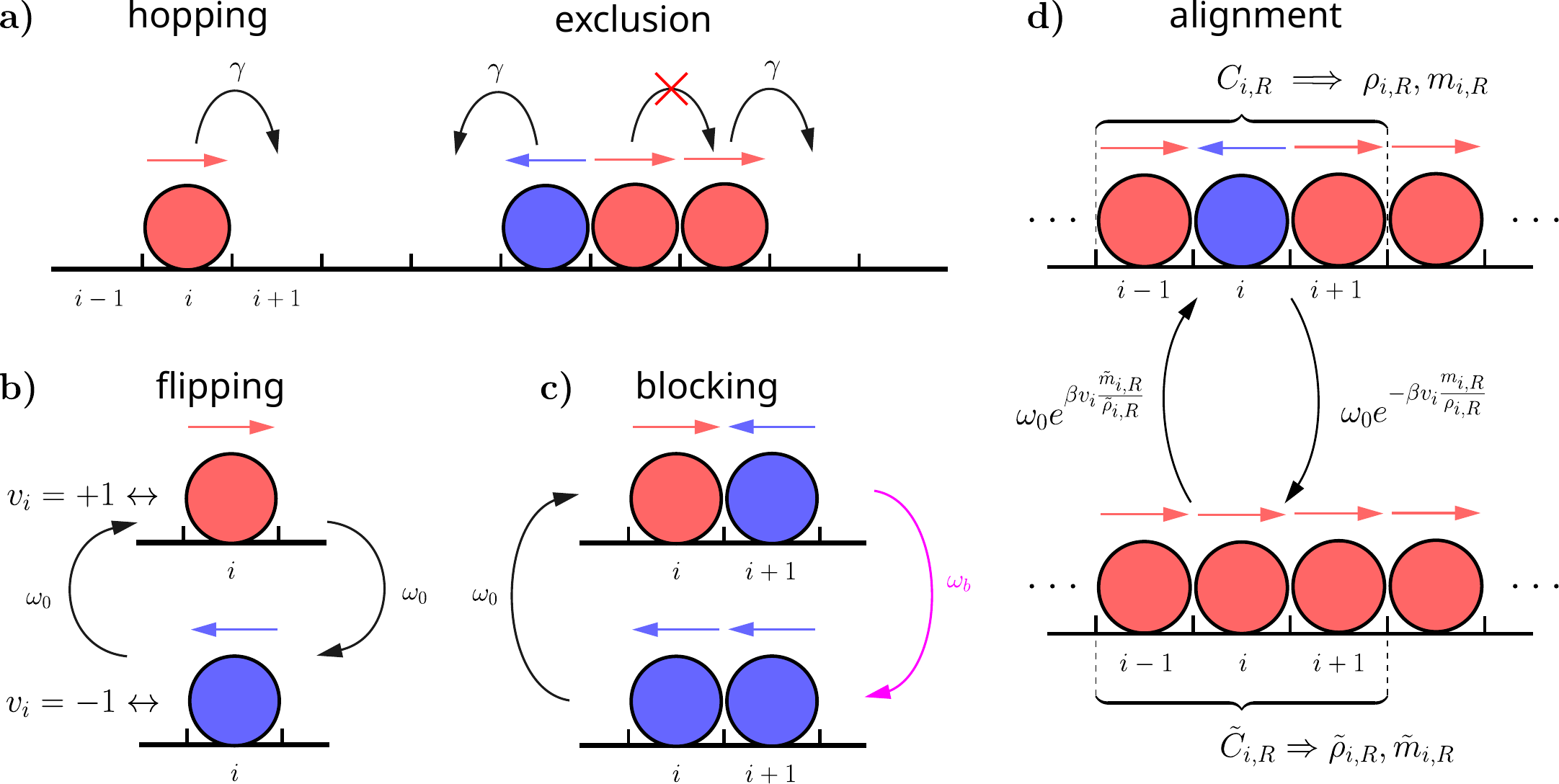}
    \caption{\label{fig_model_explanation} 
    Updating rules for the discrete stochastic lattice models.
    a) Red particles have velocity $v=1$, blue particles $v=-1$, cf.~also the color code
    in the kymographs.
    In both models, if site $i+v_i$ is empty, then particles hop to that site at rate $\gamma$. 
    If the next site is occupied, the particle cannot hop (exclusion). 
    b) Particles may flip velocity with rate $\omega_0$, if they have no nearest neighbors. 
    c) Blocking model: A particle with no neighbor in rolling direction
    flips its velocity $v_i$ at rate  $\omega_0$. 
    If site $i+v_i$ is occupied and the other particle faces 
    in the direction $-v_i$, then the particle flips with a reduced rate $\omega_b$.
    d) Alignment model: The flip rate $\omega$ depends on the 
    sub-cluster $C_{i,R}$ (particles in the same cluster having at most a distance $R$ to particle $i$) 
    containing the particle on site $i$. $\rho_{i, R}$ and $m_{i, R}$ denote the number of particles 
    and the ``polarization`` of the sub-cluster $C_{i, R}$, see appendix C for details.
    }
\end{figure}

Therefore, we introduced a second model which we call ``exclusion model with alignment''.
Here, instead of a reduced switching rate when the way is blocked,
we use an approach inspired by the 
Active Ising Model (AIM) \cite{Solon2013,Solon2015,Benvegnen2022}
that allows to implement a Vicsek-type alignment of the rolling directions.
The updating rules for this model are explained in  
Fig.~\ref{fig_model_explanation}a),b),d) and appendix C. 
In brief, the original AIM had to be changed two-fold:
first, since fibers cannot cross, excluded volume has to be implemented, as for the previous model
with blocking. Second, an alignment interaction is introduced via an AIM-type 
aligning flip rate, 
\begin{equation}\label{alignfliprate}
\omega_0 e^{-\beta v_i \frac{m_{i, R}}{\rho_{i, R}}}
\end{equation}
depending on the cluster structure, i.e.~the particle number in the cluster, $\rho_{i, R}$, 
and ``polarization'', $m_{i, R}$, given by the difference in rolling directions;
see appendix C for details.
The model introduces two new parameters, 
an aligning interaction ``radius'' $R$, inspired by the Viscek model\cite{Vicsek},
and $\beta$, inspired by the AIM, corresponding to the ``alignment strength''. 
Since fiber alignment happens upon contact,  
the aligning interaction acts only on particles contained 
in the same "cluster" (cf.~our definition of a cluster discussed in appendix A). 

\begin{figure}[t!]
    \centering
    \includegraphics[width=.99\linewidth]{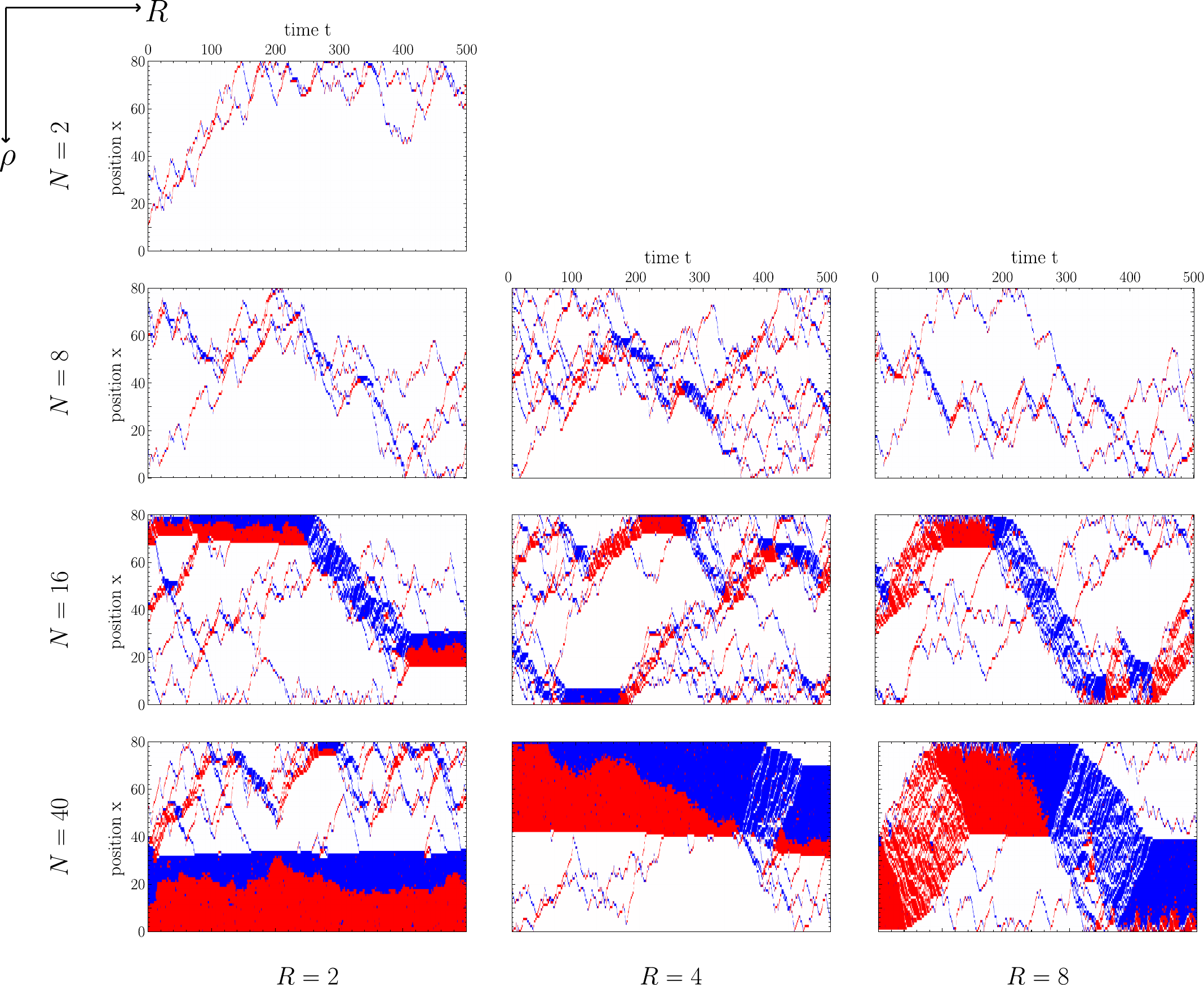}
    \caption{\label{figVRTkymosbeta3periodicbc}
    Numerically obtained kymographs of the ``exclusion model with alignment'' 
    for different particle numbers $N$ (and hence densities) and
    alignment radii $R$. The color code is the same as in the previous figures.
    While doublets are absent, the kymographs for $N\geq2$ and moderate $R$
    look qualitatively as in the experiment. 
    Other simulation parameters: 
    $\gamma=1$,
    $\omega_0=0.28$, $\beta=3$, $L=80$ (in units of fiber diamters), $t=500$.
    Note the reflective boundary conditions.
    }
\end{figure}

Numerically obtained kymographs for the  model with alignment, for collectives with $N\geq2$,
are shown in Fig.~\ref{figVRTkymosbeta3periodicbc}. 
While not recovering the doublet state (cf.~$N=2$, upper panel), 
this model yields very reasonable kymographs for larger $N$ 
and moderate alignment radius $R$ (cf.~$N=8, 16$ and $R=4$).  
This is also reflected in that
this model performs much better in describing
the collective properties, as shown Fig.~\ref{fig_model_results}b)-d).

%

Fig.~\ref{fig_model_results} shows a quantitative comparison between
the experimental data and the models. In the numerics, cf.~appendices B and C,
space and time units are the lattice spacing (which corresponds to the size of 
the particles) and the hop-rate $\gamma$ (i.e.~the particle velocity), which both are set to $1$.
For comparison, the experimental data was rescaled to these ``natural'' units,
i.e.~length was rescaled in terms of the fiber diameter of $0.6{\rm mm}$ 
and time such that the velocity peaks in Fig.~\ref{fig_velocity_distribution_flat_curved}b) 
are located at $1$.
In all panels, the blue and red curves correspond to the two experimental
temperatures. Fig.~\ref{fig_model_results}a) shows a comparison of the 
experimental cluster size distribution 
to the model with blocking,
with the numerical curves color coded by
the ratio of the free flip rate over the blocked flip rate, $\omega_0/\omega_b$.
As one can see, decreasing the blocking rate has not much influence,
the probability of larger clusters gets only slightly increased. 
That is why, while nicely capturing the ``doublets'' shown in Fig.~\ref{fig_doublet},
the blocking model was rejected for further quantitative analysis.

A similar comparison of the 
experimental cluster size distribution, now 
to the model with alignment, is shown in Fig.~\ref{fig_model_results}b).
The numerical curves are now color coded by
the alignment strength $\beta$ and dotted/dashed lines reflect small/large
alignment radius $R$.
As can be seen, the experiment at the colder temperature
($T = 150^\circ\text{C}$, blue solid curve) can be recovered by a low $\beta\approx1$,
while the one with the hotter temperature
($175^\circ\text{C}$, red solid curve) by an increased value of $\beta\approx3$.
The model with alignment hence can describe both the occurrence of larger clusters
and the temperature dependence rather well.

\begin{figure}[t!]
	\centering
	\includegraphics[width=0.99\linewidth]{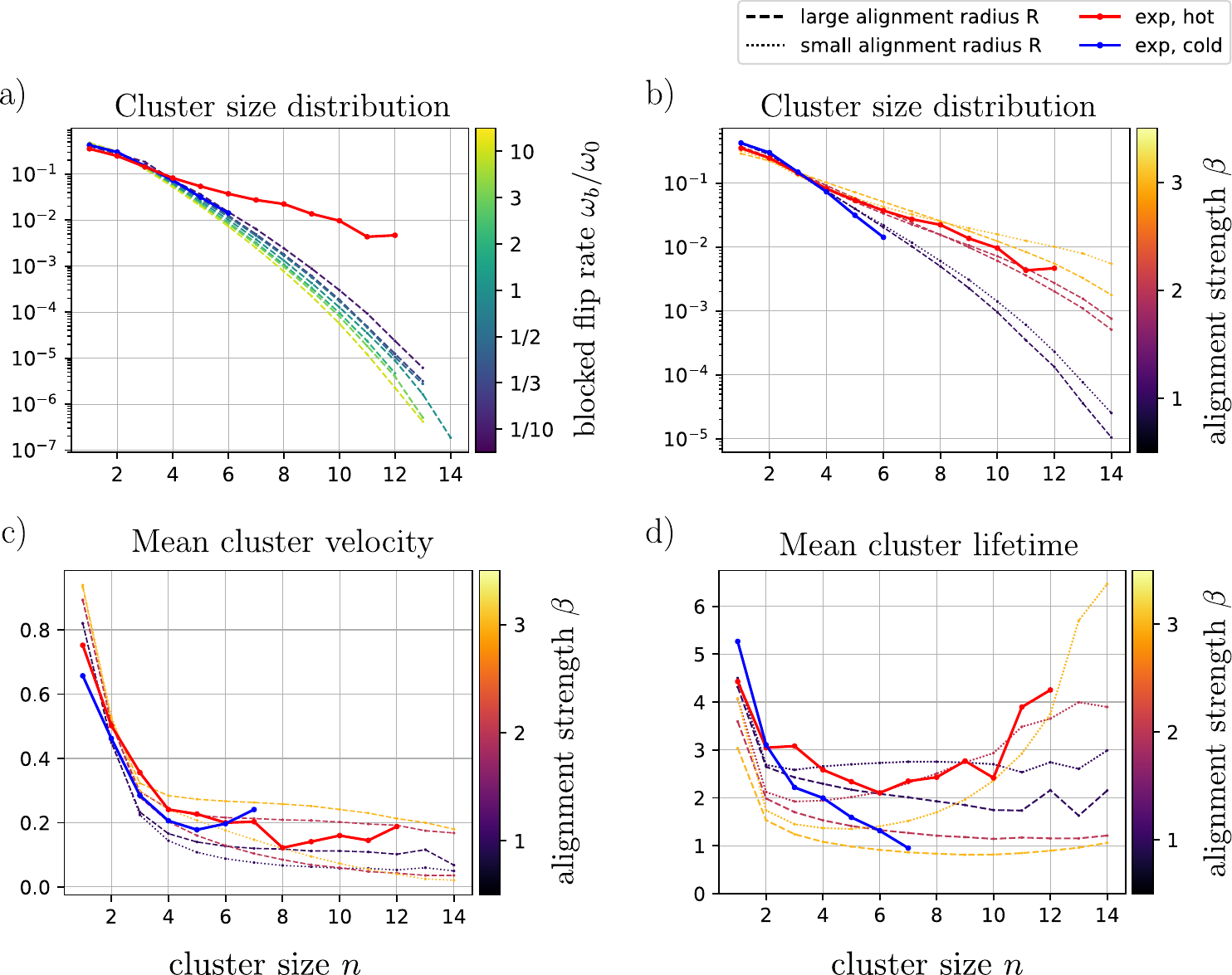}
	\caption{\label{fig_model_results}
	Comparison of experimental and numerical data. 
	a) and b) show a comparison of the experimental cluster lifetime distribution 
	($T = 150^\circ\text{C}$ as blue solid curve, $175^\circ\text{C}$ as red solid curve)
	to the exclusion model with 
	blocking (a) and with alignment (b), respectively. 
	The numerical results are color-coded in a) by 
     the ratio of the free flip rate over the blocked flip rate, $\omega_0/\omega_b$
     and in b) by the alignment parameter $\beta$ with dotted/dashed lines reflecting 
     small/large alignment radius $R$ ($R=2$ vs.~$R=32$).
	Panel c) shows the mean cluster velocity 
	and d) the mean cluster lifetimes, both for the exclusion model with alignment. 
	Data was rescaled such that 
	fiber diameter is $1$ and fiber velocity $v=1$.	
	}
\end{figure}

Fig.~\ref{fig_model_results}c) and d) show a comparison
of the other measured quantities,
the mean cluster velocity and the mean cluster lifetime,
to the model with alignment.
For the mean cluster velocity, both the experiment and the model show 
a rapid drop of velocity until $n=3$, followed by a much shallower 
decay, meaning that larger clusters move with similar velocity. 
The cluster velocities do not depend much on temperature,
which is qualitatively reflected by the weak dependence on $\beta$.
Finally, the mean cluster lifetime decreases with cluster size for the cold
experimental temperature. This effect is less pronounced for the
hotter temperature and in this case, larger clusters have an increased
lifetime again.
This is probably an effect of the finite size of the system: 
the lifetime is determined by fibers leaving and by fibers entering the cluster.
If a large cluster exists, there are only few single fibers left that could enter,
while the probability of a fiber at the
boundary leaving a cluster is rather independent of cluster size. 
Hence the lifetime is effectively larger for larger cluster,
an effect that can be captured for large $\beta$.

Overall, the model with alignment is reproducing the experiments
quite well, indicating that alignment is a main player in the collective dynamics. 
The most important parameter to describe the behavior is the alignment strength $\beta$. 
By the comparison
to the experimental data shown in Fig.~\ref{fig_model_results},
$\beta$ has to be larger for the more active system at higher
temperature. In fact, as $\beta$ enters the Boltzmann factor 
in Eq.~(\ref{alignfliprate}), the switch rate is reduced for larger $\beta$
(and $v_i$ and $m_{i,R}$ having the same sign)
rendering clusters more persistent. 
The alignment radius $R$ seems not to have a large effect,
hinting at nearest or next nearest neighbor alignment 
being dominating in the experiment. A mechanism for this alignment
could be the fibers' slight curvature or possible hydrodynamic coupling
through the cooling air flow in the environment, as discussed below.

\section{Discussion and outlook}

We have explored a novel mode of fiberboid motion that we call twitching. 
It emerges on curved substrates that due to geometrical constrains 
and motion transients tend to induce intermittent motion with frequent, stochastic 
direction reversals. Under these conditions, we could explore the emergence of collective 
behavior in an active one-dimensional gas by a simple table-top experiment, 
starting from two fibers going to the collective of many.

When two twitching fibers engage head on while running in opposite directions,
they cannot bypass each other. They display a surprising change of behavior 
and their random twitching transforms into persistent, "stubborn goat behavior", 
similar to two goats on a bridge pushing against each other, see Fig.~\ref{fig_doublet}.
This results in a ``blocked stalemate'' situation that can last for extended periods of time
before one of the fibers randomly flips direction and eventually aligns its velocity 
with the second one. The emerging persistence of the stubborn goats could be 
attributable to the two fibers rolling and pushing down against each other
such that the necessary intermediate process for direction reversal, likely an out-of-plane event, 
is effectively suppressed by the tangential force pressing on each fibers contact point 
in the negative z direction.

\begin{figure}[t!]
	\centering
	\includegraphics[width=0.99\linewidth]{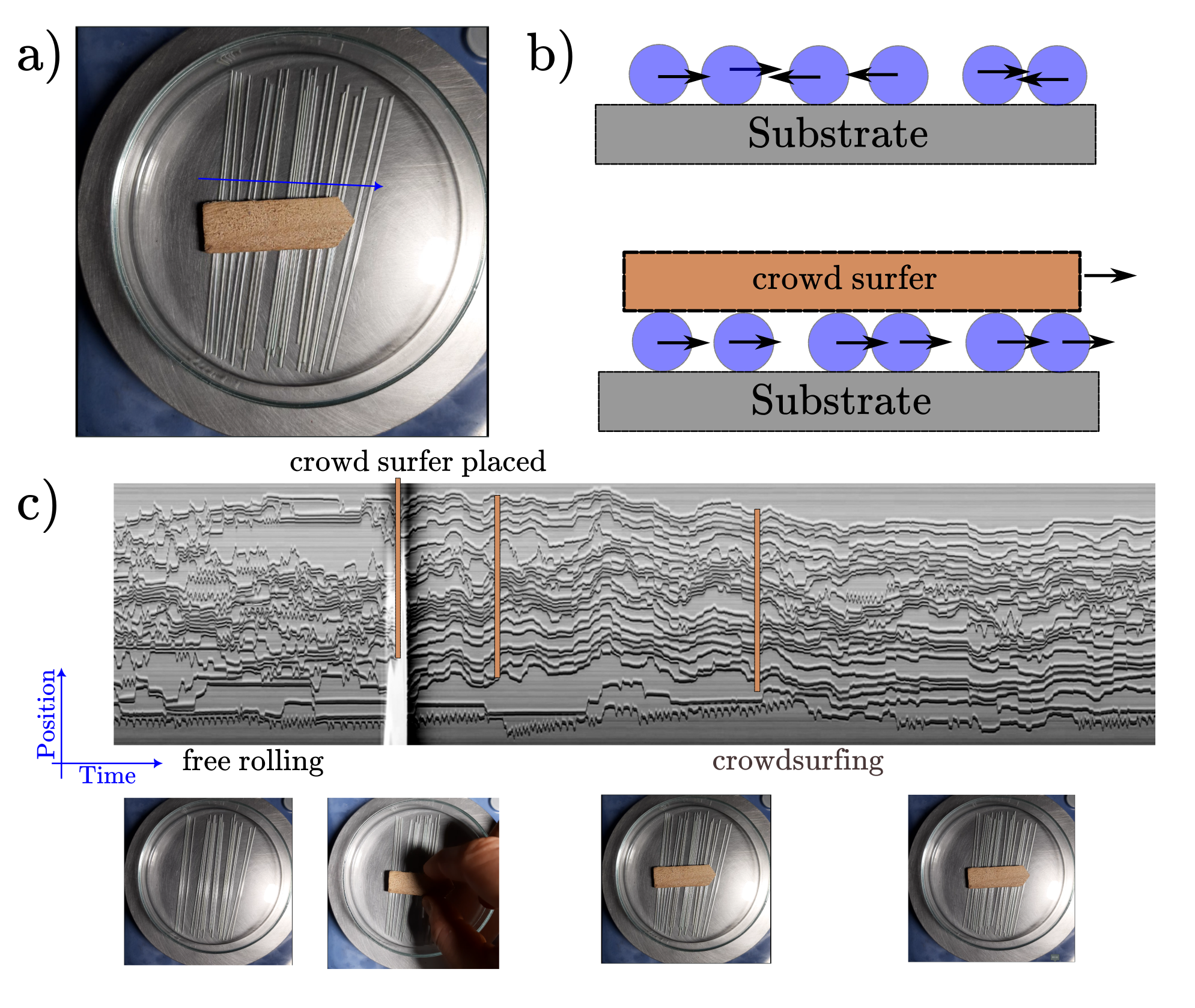}
	\caption{\label{fig_crowd-surfer}
	Crowd-surfing passive particles. 
	a)-b) Placing a large passive cargo (a wooden chip, brown) as a second layer 
	on top of twitching fiberboids increases the collective motion of clusters.    
	c) A kymograph of the fiber motion together with snapshots at different 
	time points (x-axis of kymograph). Upon first contact with the crowd surfer 
	(white strip in kymograph) the cluster dynamics becomes more collective. 
	Fibers are running at $T=150^\circ C$, total time span of the kymograph is 36s.  
	See supplementary movie 7.
	}
\end{figure}

The experimental observation of two stubborn fibers blocking each other is 
well captured by a simple ``exclusion model with blocking". 
However, switching from a pair of fibers to three and more fibers in the collective 
again radically changes the behavior. The blocking behavior seen for isolated fiber 
pairs is no more observable while ordered clusters running at non-zero speeds 
in either of the two directions become the predominant dynamic structures. 
We have shown that the formation of such collective clusters can be consistently 
modeled with an additional (finite range) interaction, aligning the velocities 
of fibers within the same cluster. 
We speculate here that the physical origin of the alignment interaction could be 
related to one of the following phenomena: 
a) the slight in-plane curvature of fibers (rolling effectively as bent arcs),  
that tends to favor denser cluster packing of fibers running in the same direction 
within the cluster and
b) a hydrodynamic coupling of the cooling air flow 
on top of fiber clusters, biasing the dynamics of individuals.
Both would tend to favor coherent behavior of clusters.
The presence of an extended hydrodynamic medium to which the fibers and their clusters 
couple and the related hydrodynamic power-law dynamics associated with turbulent 
phenomena in the advected/cooling surrounding air flow resonate more with the 
observed power-law tails of the cluster lifetimes, cf.~Fig.~\ref{fig_cluster_life}, 
and would somehow favor explanation b). However, the presence of other power-law 
decaying collective modes in clusters including internal density waves could similarly 
account for some form of power-law behavior.
While the bets, which of the two mechanisms (if any) are at play are still open, 
the change of behavior from individual, over binary to collective fiber dynamics
remains one of the interesting experimental challenges in the surprisingly complex
system of twitching fiberboids and their collectives.

Another interesting feature of twitching fiberboids is their ability to switch behavior 
to persistent rolling depending on the presence of other passive particles 
that enhance their planar confinement. For a single twitching fiberboid, this was 
shown in Fig.~\ref{fig_AluFoil-Ball}. However, one can also use larger objects
that effectively couple several fiberboids, as shown in Fig.~\ref{fig_crowd-surfer}
and supplementary movie 7,
allowing the object to ``crowd surf''. 
Both examples show that fiberboids can be effectively turned 
into active agents able of responding to their environment 
and performing individual as well collective tasks. This aspect adds an additional layer 
of complexity bearing possibly new physics, akin to  behavioral dynamics of biological agents 
from the animal kingdom like cargo carrying ants\cite{Ants}, 
spontaneously self-organizing 
to avoid a tug of war.



\section*{Appendix}

\subsection*{A\quad Cluster definition}

We define a cluster as a collection of fibers, such that inside it,
the distance to nearest neighbors is less than some maximum distance $d_\text{max}$. 
Experimentally, fibers inside a cluster are not always equidistantly placed,
and fluctuations of positions inside a cluster are visible. 
Nevertheless, a cluster can be discerned as a single moving object for substantial times. 
So, instead of setting $d_\text{max}$ equal to the thickness of the fibers themselves, 
we opted  for double this value, $2d_\text{max}$. This gives some ``slack'' and
also allows for easier comparison to the lattice models, 
where allowing for gaps between fibers allows the clusters 
to be more mobile.

In order to talk about the properties of a cluster over time, 
one needs a way to compare clusters at different times. 
We define two clusters $C(t)$ and $C^\prime(t^\prime)$ measured 
at two times $t < t^\prime$ to be the same if 
(i) they contain the exact same fibers and 
(ii) this was the case during the entire time period between $t$ and $t^\prime$. 
Our definition thus implies, that if a cluster loses 
(distance larger than $2d_\text{max}$) or gains a fiber, 
even only for some short time, than that cluster is considered dissolved
and counted as a new cluster of size $n\mp1$.

\subsection*{B\quad Details for the exclusion model with blocking}

Here we give details on the model description sketched in Fig.~\ref{fig_model_explanation}. 
The model with blocking consists of $N$ particles, placed on a lattice of size $L$ and 
lattice constant $a$. We enforce volume exclusion, i.e.~no more than one particle 
may occupy a given lattice site.
A particle located at site $x_i$ hops with rate $\gamma$ to the neighboring 
site $x_i + v_i dt$, where $dt$ is some small time step and $v_i = \pm a\gamma$ 
is the velocity of a particle. 
By redefining space and time units, we may set $\gamma=1$ and $a=1$ and thus have $v_i = \pm 1$, 
such that we can just write $x_i + v_i$ for the target site of a hop-event.  
The move is rejected, if this target site is already occupied (volume exclusion) or if it lies beyond the system domain sites $-1$ and $L$ (reflective b.c.), 
in the latter case, the moving direction of the particle is also flipped to $-v_i$.
Particles will also flip their moving direction with rate $\omega_0$ or $\omega_b$ as follows:
If the particle $i$ is ``blocked", i.e.~there is another particle $j$ 
located at site $x_i + v_i$ with opposite direction of movement $v_j = -v_i$, 
then $v_i$ flips with rate $\omega_b$, otherwise it flips with $\omega_0$, 
to which we refer to as the ``free" case. 

The model is implemented using a discrete time Monte Carlo algorithm: 
(1) Uniformly pick a particle $i$. 
(2) Perform one of the following moves:
\begin{center}
	\begin{tabular}{l r}
		Prob(hop to  $x_i + v_i$, if allowed): & $\Delta t$\\
		Prob(free: spin flips): & $\omega_0 \Delta t$ \\
		Prob(blocked: spin flips): & $  \omega_b \Delta t$\\
		Prob(no update): & $1 - (1 + \max(\omega_0,  \omega_b))\Delta t$
	\end{tabular}
\end{center}
The move is selected via tower sampling. 
(3) After $N$ updates, increase the current time by $\Delta t$, defined by 
\begin{equation}
	\Delta t  = (1 + \max(\omega_0, \omega_b))^{-1}
\end{equation}
This choice ensures that probabilities are bounded by $0$ and $1$, 
and minimizes the probability of no update occurring. 

\subsection*{C\quad Details for the exclusion model with alignment}

We again refer to Fig.~\ref{fig_model_explanation} for a sketch of the model. 
As described in the main text, the model with alignment is a modification of 
the one introduced in appendix B. 
The exclusion model with alignment is mostly identical to the exclusion model 
with blocking discussed above, except that the ``blocked" flip rate $\omega_b$ 
is replaced by a flip-rate, which induces the local alignment of particles.

Let again $x_i$ and $v_i$ denote the position and movement of direction of particle $i$. 
Now we consider they are contained in some cluster $C$ with $d_\text{max} = 2$ (c.f.~appendix A). 
To implement local interactions, we introduce a sub-cluster of size $C_{i, R}$, containing 
all particles $j$ inside $C$, such that $\vert x_i - x_j\vert < R$.
In similar notation to \cite{Benvegnen2022}, we denote by $n_{i, R}^\pm$ the 
number of particles inside $C_{i, R}$ with movement of direction $v_i = \pm 1$ respectively. 
We then define $\rho_{i, R} = n_{i, R}^+ + n_{i, R}^-$ and $m_{i, R} = n_{i, R}^+ - n_{i, R}^-$ 
to be the total number of particles and the 
total ``magnetization`` or ``polarization`` of the sub-cluster $C_{i, R}$. 
The flip rate of particle $i$ contained in the sub-cluster $C_{i, R}$ is then given by
\begin{equation}
		W(\text{clustered}: v_i \to -v_i) = \omega_0 e^{-\beta v_i \frac{m_{i, R}}{\rho_{i, R}}}
\end{equation} 
where $\omega_0$ is the free flip rate (of a single particle) 
and $\beta$ is a parameter that controls the alignment strength, 
loosely connected to an inverse temperature \cite{Solon2013,Solon2015} of algorithmic nature.
Note that this flip-rate is large if the majority of particles inside the sub-cluster 
face in the opposite direction of $i$ and small otherwise, thereby 
implementing sub-cluster alignment.

The simulation is performed analogously to the model with blocking in appendix B: 
(1) Uniformly pick a particle $i$ and compute its sub-cluster $C_{i, R}$. 
(2) Perform one the following moves with the given probabilities
\begin{center}
	\begin{tabular}{l r}
		Prob(hop to  $x_i + v_i$, if allowed): & $\Delta t$\\
		Prob(free: spin flips): & $\omega_0 \Delta t$ \\
		Prob(clustered: spin flips): & $  \omega_0 e^{-\beta v_i\frac{m_{i, R}}{\rho_{i, R}}}\Delta t$\\
		Prob(no update): & $1 - (1 + \omega_0\max(1,e^{\beta}))\Delta t$ 
	\end{tabular}
\end{center}
(3) After $N$ steps increase the time by $\Delta t$, given by
\begin{equation}
	\Delta t  = (1 + \omega_0\max(1, e^{\beta}))^{-1}
\end{equation}

\section*{Author Contributions}
F.Z.~and I.M.K. conceived the project. A.A. and I.M.K designed the experimental setup and performed experiments.  A.A. and F.Z. developed the simulation framework. A.A. performed the simulations and  experimental analysis. All authors wrote the paper.

\section*{Conflicts of interest}
There are no conflicts to declare.

\section*{Acknowledgements}
IMK thanks the members of the Theory of Polymers team at IPF Leibniz for continuous support and hospitality. He acknowledges funding by DFG-ANR project Rodrolls-490954343, ANR-21-CE30-005.



\balance


\bibliography{bib_for_ZEEM_Alex.bib} 
\bibliographystyle{rsc} 

\end{document}